\documentclass[a4paper,11pt]{article}

\usepackage{pos}
\usepackage{lipsum} 
\usepackage{lineno}

\title{Extending SkyLLH software for neutrino point source analyses with 10 years of IceCube public data}

\ShortTitle{SkyLLH extension for 10 years of public data}

\author{The IceCube Collaboration \\{\normalsize \normalfont(a complete list of authors can be found at the end of the proceedings)}\\}

\emailAdd{chiara.bellenghi@tum.de}
\emailAdd{martina.karl@tum.de}
\emailAdd{martin.wolf@tum.de}

\abstract{

Searching for the sources of high-energy cosmic particles requires sophisticated analysis techniques, frequently involving hypothesis tests with unbinned log-likelihood (LLH) functions. SkyLLH is an open-source, Python-based software tool to build these LLH functions and perform likelihood-ratio tests. We present a new easy-to-use and modular extension of SkyLLH that allows the user to perform neutrino point source searches in the entire sky using ten years of IceCube public data. To guide the user, SkyLLH provides tutorials showing how to analyze the experimental data and calculate useful statistical quantities. Here we describe the details of the analysis workflow and illustrate some of the possible methods to work with the IceCube public dataset. Additionally, we show that SkyLLH can reproduce the results from a previous IceCube publication that used the public data release. We obtain a similar local significance for the neutrino emission from a list of candidate sources within a maximum shift of 0.5$\sigma$. Finally, the measured neutrino flux from the most significant source candidate, NGC~1068, shows substantial agreement with the previously published result.

\vspace{4mm}
{\bfseries Corresponding authors:}
Chiara Bellenghi$^{1*}$, Martina Karl$^{1,2}$, Martin Wolf$^{1}$\\
{$^{1}$ \itshape Technical University of Munich, TUM School of Natural Sciences, Department of Physics, James-Franck-Straße 1, D-85748 Garching bei M\"unchen, Germany}\\
{$^{2}$ \itshape European Southern Observatory, Karl-Schwarzschild-Straße 2, D-85748 Garching bei M\"unchen, Germany}\\[4mm]
$^*$ Presenter

\ConferenceLogo{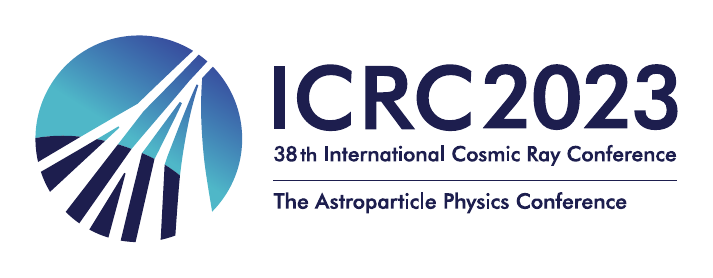}

\FullConference{The 38th International Cosmic Ray Conference (ICRC2023)\\ 26 July -- 3 August, 2023\\ Nagoya, Japan}
}

\graphicspath{{./}{figures/}}

\begin{document}

\maketitle

\section{Introduction}
\label{sec:introduction}
\noindent SkyLLH \cite{Kontrimas:2021icrc} is an open-source, Python-based, and easy-to-use software framework to build log-likelihood (LLH) functions for celestial event data. It has been developed within the IceCube collaboration \cite{Aartsen:2016nxy} to perform frequentist statistical data analyses for point-like neutrino source searches, including the one that resulted in a  $4.2\,\sigma$ evidence for neutrino emission from the Seyfert II galaxy NGC~1068 \cite{doi:10.1126/science.abg3395}. 
In the past, IceCube performed searches for steady and flaring neutrino point-like sources using 10 years of point-source data \cite{PhysRevLett.124.051103, IC-2021-10year-multi-flare}. This data has been made public \cite{IceCube:2021xar}.
In order to make this information accessible and usable for the entire community, we present an extension to the SkyLLH software framework which now includes an interface to use this latest all-sky point-source data release. In the following, we briefly describe the content of the data release and detail the likelihood method that has been implemented for point-like neutrino searches.
The published experimental data are provided together with binned detector response functions that describe how a muon produced in the interaction of a detectable neutrino would be reconstructed in IceCube.
Among the reconstructed muon observables given in the detector response functions is the angular separation between the parent neutrino and the reconstructed muon. This quantity is fundamental for inferring whether the spatial distribution of the events in the sky is consistent with emission from a point-like source. On the other hand, the reconstructed muon direction is also required in the formulation of the likelihood function used in Ref.~\cite{PhysRevLett.124.051103} but is not provided in the detector response functions. Nevertheless, we can address this limitation by using a slightly modified version of the likelihood. We investigate the effect of this modification by comparing the likelihoods of the internal and the public analyses. Furthermore, we fit the source candidates that were searched for neutrino emission in Ref.~\cite{PhysRevLett.124.051103} and compare their significance to the published one. For NGC~1068 specifically, we measure the neutrino flux and compare the result to the one reported in Ref.~\cite{PhysRevLett.124.051103}. Finally, we briefly describe a method to fit time-dependent neutrino emission profiles from point-like sources, which is implemented in the software interface for public neutrino data analyses described in this work.

\section{The 10 years of public neutrino point-source data}
\label{sec:data_release}
\noindent In January 2021, IceCube released ten years of data for neutrino point-source searches \cite{IceCube:2021xar}. The data spans the time range from April 6, 2008 to July 10, 2018, covering five different data acquisition periods (seasons) corresponding to different detector configurations: IC40, IC59, IC79, IC86 2011, and IC86 2012--2017. For details regarding the IceCube detector see Ref.~\cite{Aartsen:2016nxy}. The data sample is optimized for the identification of neutrino clusters, which requires good angular resolution. Therefore, it includes track-like events from all directions in the sky. In the Southern sky, the data sample is dominated by cosmic-ray-induced atmospheric muons, which are instead absorbed in the other hemisphere when crossing the Earth. Therefore, events from the Northern sky are mostly produced in neutrino interactions.
The data contains 1,134,450 recorded events with their best-fit reconstructed muon observables $\vec{x} = (E_{\mu}, \vec{d}_{\mu}, \sigma_{\mu}, t_{\rm{obs}})$: the energy $E_{\mu}$, direction in right-ascension $\alpha$ and declination $\delta$, $\vec{d}_{\mu} = (\alpha_{\mu}, \delta_{\mu})$, the uncertainty on the reconstructed muon direction $\sigma_{\mu}$, as well as the observation time $t_{\rm{obs}}$. It should be noted that the angular uncertainty includes a correction for the kinematic angle between the parent neutrino and the reconstructed muon, important for neutrino energies $\lesssim1$~TeV.
In addition, the release provides tabulated instrument response functions (IRFs) consisting of muon-neutrino effective areas, $A_{\mathrm{eff},\nu_{\mu}+\overline{\nu}_{\mu}}$, and binned probabilities in the form $M(E_\mu,\Psi_{\nu\mu},\sigma_{\mu}|\delta_\nu,E_\nu)$. Thus, for a given parent neutrino with declination $\delta_\nu$ and energy $E_\nu$, $M$ contains the probability of reconstructing a neutrino-induced muon with $E_\mu$, $\sigma_\mu$, and an angular separation $\Psi_{\nu\mu}$ between the parent neutrino and the muon. $M$ is referred to as the detector response matrix from here on\footnote{It is denoted as \textit{smearing matrix} in the data release \cite{IceCube:2021xar}.}.
The published IRFs allow the construction of the probability density functions (PDFs) required in a hypothesis log-likelihood ratio test.

\section{Likelihood formalism}
\label{sec:lh-formalism}
\noindent The IceCube point-source searches look for spatial and/or temporal clusters of astrophysical neutrinos over the atmospheric muon and neutrino, and diffuse cosmic neutrino backgrounds \cite{IceCubeDiffuse2022ApJ}. 
To be able to discriminate between signal and background events, IceCube uses an unbinned likelihood-ratio hypothesis test, where two hypotheses are fitted to the data, and their likelihoods are compared:
\begin{itemize}
    \item $H_0$: the observed data only consists of background muon events, induced by atmospheric muons and neutrinos or by the astrophysical diffuse neutrino flux;
    \item $H_1$: the observed data consists of, both, background and signal muon events, where the latter are induced by the interaction of astrophysical neutrinos originating from a point-like neutrino source, emitting with strength proportional to the mean number of signal events in the detector, $n_{\mathrm{s}}$, and following a power-law energy spectrum $\propto E^{-\gamma}$.
\end{itemize}
The two hypotheses are nested. Hence, for $n_{\mathrm{s}} \equiv 0$, the signal hypothesis, $H_1$, reduces to the background-only one, $H_0$.
The maximum log-likelihood ratio is then the test-statistic, TS, used for the analysis:
\begin{equation}\label{eq:TS}
    \mathrm{TS} = -2\log{\frac{\sup{_{H_0}}\left[ \mathcal{L}(H_0|\vec{x})\right] }{\sup_{H_1}\left[ \mathcal{L}(H_1|\vec{x}) \right]}},
\end{equation}
where $\mathcal{L}(H|\vec{x})$ is the likelihood function of $H_0$ or $H_1$, given the recorded data $\vec{x}$. The likelihood function is commonly represented as a mixture model, consisting of the weighted sum of a signal PDF and a background PDF, $\mathcal{S(\cdot)}$ and $\mathcal{B}(\cdot)$, respectively \cite{BRAUN2008299}:
\begin{equation}
    \mathcal{L}(n_{\rm{s}}, \gamma|\vec{x})= \prod_{i=1}^{N}\left[\frac{n_s}{N}\mathcal{S}(x_i|\gamma) + \left(1-\frac{n_s}{N}\right)\mathcal{B}(x_i)\right],
\label{eq:LH}
\end{equation}
where $N$ is the total number of events $x_i$ in the sample.

\noindent For the nested hypotheses used in point-source searches as described here, the null hypothesis $H_0$ has no free parameters, hence the test-statistic can be written as a simplified expression containing the signal-over-background PDF ratio:
\begin{equation}
    \mathrm{TS} = 2\sum_{i=1}^{N} \log\left[ \frac{\hat{n}_{\rm{s}}}{N}\left( \frac{\mathcal{S}(x_i|\hat{\gamma})}{\mathcal{B}(x_i)} - 1 \right) + 1 \right].
\label{eq:ts-llh-ratio}
\end{equation}
Here the parameters $\hat{n}_{\mathrm{s}}$ and $\hat{\gamma}$ denote the values of the two parameters that maximize the likelihood in Eq.~(\ref{eq:LH}).

\noindent The new extension of SkyLLH for the public data provides classes and functions to construct the signal and background PDFs using the provided experimental data and IRFs. By using the law of conditional probability, both the signal and background PDFs can be separated into a spatial and an energy part.
Spatial and energy background PDFs are defined identically to the internal traditional IceCube analysis (see \textit{e.g.} Ref.~\cite{PhysRevLett.124.051103}), therefore we assume that the background flux is uniformly distributed in space and that its energy distribution can be approximated by the energy distribution of the experimental data. This assumes that the experimental data is strongly dominated by background events, which is a safe assumption in this case.

\noindent The signal PDF for a point-like source with a power-law energy spectrum is dependent on the source declination $\delta_{\mathrm{src}}$\footnote{Due to the location of IceCube at the geographic South Pole, the distributions of the recorded events are symmetric in azimuth and hence depend in good approximation solely on the declination of the event.} and the spectral index $\gamma$ and can be expressed as
\begin{equation}
    \mathcal{S}(\delta_i, E_i|\delta_{\rm{src}},\gamma) = \mathcal{S}_{\rm{spatial}} \cdot \mathcal{S}_{\rm{energy}} = P_\mathrm{PSF}(\delta_i|\delta_{\mathrm{src}}) \cdot P_{\rm{E}}(E_i|\delta_{\mathrm{src}},\gamma),
    \label{eq:signal-pdf}
\end{equation}
where $P_{\mathrm{PSF}}$ is the point-spread-function (PSF) of the detector for a given source position and $P_{\rm{E}}(E_i|\delta_{\mathrm{src}},\gamma)$ is the signal energy PDF to observe an event with energy $E_i$ assuming point-like neutrino emission from $\delta_{\mathrm{src}}$ and spectral index $\gamma$. For the PSF an analytical function in the form of a symmetric two-dimensional Gaussian is used (\textit{e.g.} Ref.~\cite{BRAUN2008299}, and Ref.~\cite{PhysRevLett.124.051103}).

\noindent The signal energy PDF can be constructed from the detector response matrix $M$ and can be formulated as an integral over all parent neutrino energies: 
\begin{equation}\label{eq:energy_pdf}
    P_{\rm{E}}(E_i|\delta_{\mathrm{src}},\gamma) \equiv P(E_{\mu}|\delta_{\nu},\gamma) = \int_{E_{\mathrm{min}}}^{E_{\mathrm{max}}} dE_{\nu}\,P(E_{\mu}|E_{\nu},\delta_{\nu})\,P(E_{\nu}|\gamma)\,P(E_{\nu}|\delta_{\nu}).
\end{equation}
The integrand is the product of three probabilities. First, $P(E_{\mu}|E_{\nu},\delta_{\nu})$ is the probability of $E_\mu$ given the parent neutrino energy $E_\nu$ and declination $\delta_\nu$, which is calculated from the detector response matrix $M$. Second, the probability of the parent neutrino energy for the given power-law energy spectrum with spectral index $\gamma$, denoted as $P(E_{\nu}|\gamma)$. Third, the probability of detecting the parent neutrino energy given its direction, $P(E_{\nu}|\delta_{\nu})$, is calculated from the provided tabulated effective areas. The integration limits $E_{\rm{min}}$ and $E_{\rm{max}}$ are set in order to cover the parent neutrino energy range of $10^2 - 10^9$~GeV, as provided in the IRFs.

\begin{figure}
    \centering
    \includegraphics[width=.9\linewidth]{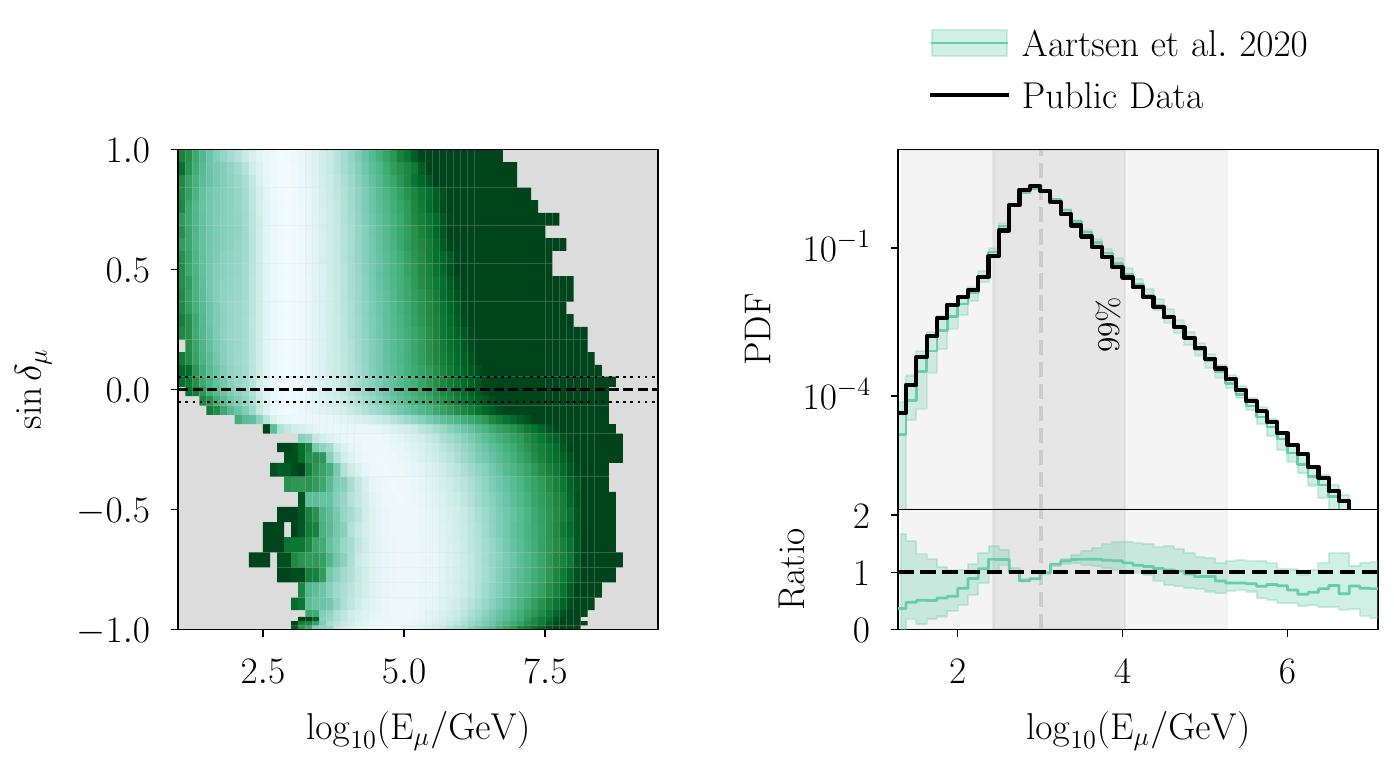}
    \caption{Comparison of the signal energy PDFs for the season IC86 2012--2017 and spectral index $\gamma=3.2$. {\bf Left}: PDF as used in Ref.~\cite{PhysRevLett.124.051103}. The lighter the color, the higher the probability density. The black lines delimit $\delta_{\mu}=(-0.01\pm3.00)^{\circ}$. {\bf Right}: 1D comparison between the internal IceCube PDF (green) at reconstructed muon declination $\delta_{\mu}=-0.01^{\circ}$ and the one used in the public data analysis (black) at source declination $\delta_{\mathrm{src}}=-0.01^{\circ}$ (NGC~1068). The shaded green band represents the variations of the internal energy PDF when considering events reconstructed at declinations enclosed by the dotted black lines in the left-hand plot. The energies of the experimental data fall in the gray shaded area, with the central 99\% quantile indicated by the darker grey area. The lower panel shows the ratio between the public PDF and the internal one for all events reconstructed within $3^{\circ}$ from the source.}
    \label{fig:sig-energy-pdfs}
\end{figure}

\noindent Only the mathematical representation of the signal energy PDF differs from the internal analysis used in Ref.~\cite{PhysRevLett.124.051103}.
Internally, it is constructed as $P(E_\mu|\sin(\delta_{\mu}),\gamma)$, whereas the public data analysis uses the signal energy PDF given in Eq.~(\ref{eq:energy_pdf}). The difference is illustrated in Figure \ref{fig:sig-energy-pdfs} for the dataset IC86 2012--2017 and for a source at the location of NGC~1068 with its best-fit spectral index $\gamma=3.2$ \cite{PhysRevLett.124.051103,doi:10.1126/science.abg3395}. The internal analysis uses the approximation $\delta_{\mathrm{src}}\approx\delta_{\mu}$ to simplify the calculation of the signal-over-background PDF ratio (see Eq.~(\ref{eq:ts-llh-ratio})), which then has the same data structure as the signal and background PDFs. With the public detector response matrix, the construction of such a signal energy PDF is impractical due to missing information about the reconstructed muon declination $\delta_\mu$.

\section{Analysis workflow with SkyLLH}
\label{sec:analysis-workflow-with-skyllh}
\noindent The SkyLLH software framework is designed to follow the mathematical structure of the log-likelihood ratio formalism in Eq.~(\ref{eq:ts-llh-ratio}). Hence, Python base classes define interfaces for PDFs, PDF ratios, log-likelihood ratio, and test-statistic functions. The \verb|Analysis| class bundles all components of an analysis into a single object.
Despite classes and functions to evaluate the log-likelihood function, SkyLLH also provides utility classes for data handling. For instance, the \verb|Dataset| class allows the definition of a data set with observed data and simulations stored on disk, and its loading into memory.
The public data interface described in this contribution creates the \verb|Analysis| object through a \verb|create_analysis| Python function, which constructs all the necessary PDFs and PDF ratios, as well as the log-likelihood function.
After the construction of the \verb|Analysis| object, the data can either be unblinded via the \verb|unblind| method or a pseudo-experiment with background and signal events can be generated via the \verb|do_trial| method of the \verb|Analysis| class.
The documentation\footnote{\url{https://icecube.github.io/skyllh/master/html/index.html}} of SkyLLH provides a tutorial to load the public data and to create the \verb|Analysis| object for a single point-source with a power-law energy flux function to reproduce the results published in Ref.~\cite{PhysRevLett.124.051103}. Furthermore, it shows how to generate pseudo-experiments for sensitivity and significance calculations.
Notably, SkyLLH's design allows for a combined data analysis of multiple data sets or detectors by taking the product of all likelihood functions, one for each individual data set. Thus, the public data could be combined with data from other experiments. 

\section{Reproducibility of published IceCube point-source search results}
\label{sec:reproducibility}
\begin{figure}
    \centering
    \includegraphics[width=.83\linewidth]{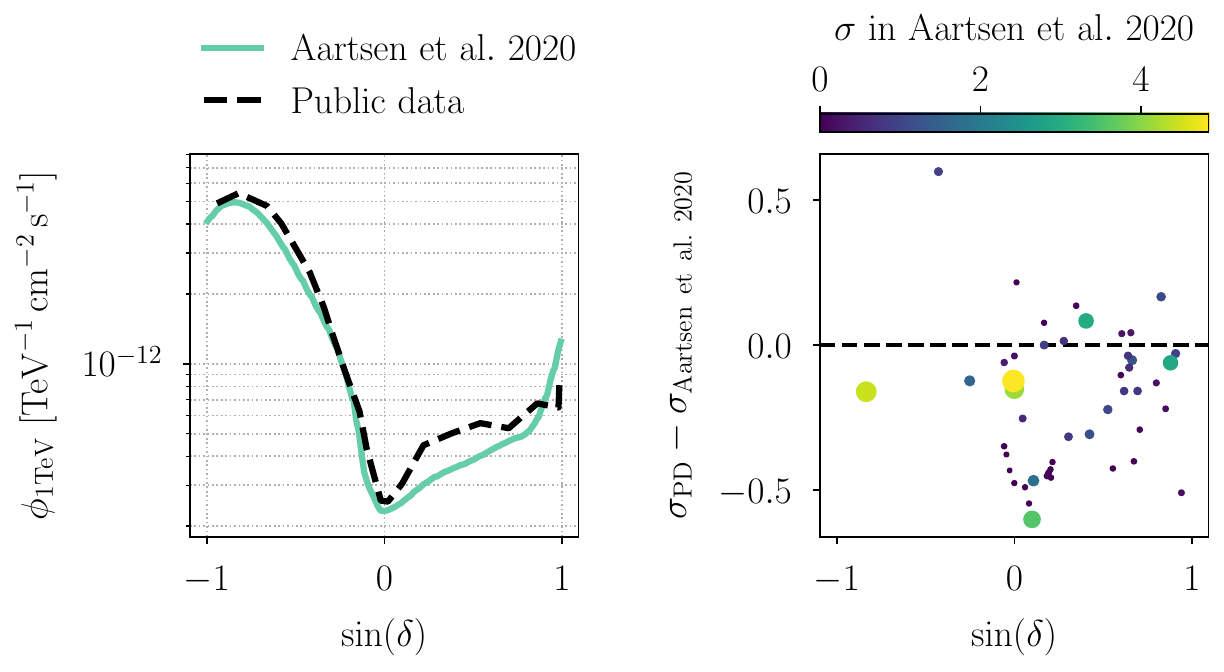}
    \caption{Reproducibility of published results. {\bf Left:} Comparison of sensitivity to an astrophysical neutrino flux as a function of the position in the sky assuming an $E^{-2}$ power-law energy spectrum for the analysis used in Ref.~\cite{PhysRevLett.124.051103} (green solid line) and the public data (PD) analysis presented in this work (dashed black line). {\bf Right:} Shift of local significance in terms of Gaussian-equivalent standard deviations as a function of the sine of the declination compared to the result published in Ref.~\cite{PhysRevLett.124.051103}. The dots represent the objects in the list of candidate sources in Ref.~\cite{PhysRevLett.124.051103} and the two most significant locations in the northern and southern sky. The color scale, as well as the size of the dots, represent the published local significance. The green point with the largest shift to lower significances (at the very bottom of the panel) corresponds to TXS~0506+056.}
    \label{fig:sensitivity-comparison}
\end{figure}

\begin{figure}
    \centering
    \includegraphics[width=.47\linewidth]{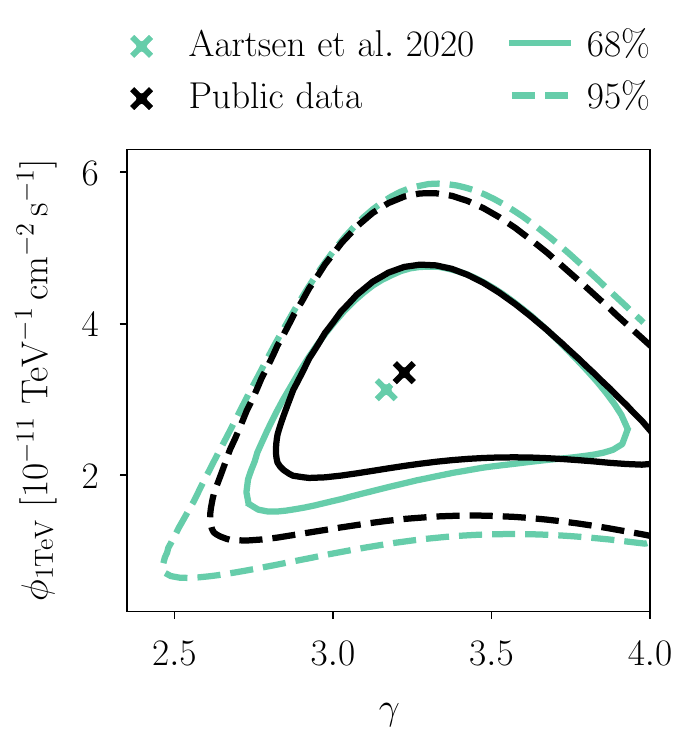}
    \caption{Comparison of the likelihood scan around the best-fit flux parameters ($\phi_{\mathrm{1TeV}},\ \gamma$) of NGC~1068. The best-fit flux (cross), $1\sigma$ (solid line), and $2\sigma$ (dashed line) confidence levels from Wilks' theorem are shown in green for the analysis in Ref.~\cite{PhysRevLett.124.051103} and in black for the analysis on public data presented in this work.}
    \label{fig:llh_scan}
\end{figure}

\noindent To test how well the analysis published in Ref.~\cite{PhysRevLett.124.051103} can be reproduced using the public data release and the software interface described in this contribution, we compare their respective sensitivities to an $E^{-2}$ flux of astrophysical neutrinos produced by a point-like source. The result of this comparison is illustrated in the left panel of Figure \ref{fig:sensitivity-comparison} where the sensitivity flux is shown as a function of the sine of the declination for the public data (black dashed) and the internal IceCube analysis (green) as published in Ref.~\cite{PhysRevLett.124.051103}. In the right panel of Figure \ref{fig:sensitivity-comparison}, the local significance difference in terms of Gaussian-equivalent standard deviations for the source candidates analyzed in Ref.~\cite{PhysRevLett.124.051103} is plotted as a function of their position in the sky. For most of the tested sources, the significance is weaker when using the public data (PD), but it is reproduced within $\sim0.5$ standard deviations.
Overall, the public data analysis is slightly less sensitive due to the coarse binning of the detector response matrix, which contains only 3 bins for the parent neutrino direction, covering the Southern sky $(-90^{\circ}<\delta_{\nu}<-10^{\circ})$, the horizon $(-10^{\circ}<\delta_{\nu}<10^{\circ})$, and the Northern sky $(10^{\circ}<\delta_{\nu}<90^{\circ})$.
We measured the neutrino flux $\phi(E_{\nu})= \phi_{\mathrm{1TeV}}\,(E_{\nu}/\mathrm{1TeV})^{-\gamma}$ from NGC~1068 with the public data analysis presented in this work and find the best-fit values for the flux parameters to be $\hat n_s=60.1$, and $\hat \gamma=3.2$. The observed mean number of signal events $\hat n_s$ can be converted into a neutrino flux normalization by integrating the effective area at the source location multiplied by the power-law energy flux over the neutrino energy range and the observation time period. We obtain $\hat\phi_{\mathrm{1TeV}}=3.3\times10^{-11}\,\rm{TeV^{-1}\,cm^{-2}\,s^{-1}}$. All these values are compatible with the ones published in Ref.~\cite{PhysRevLett.124.051103}. As an additional check, we performed a likelihood scan around the best-fit flux parameters and derived the $1\sigma$ and $2\sigma$ confidence levels from Wilks' theorem \cite{WilksTheorem}. The comparison of the likelihood landscapes is illustrated in Figure \ref{fig:llh_scan} showing that the two results largely overlap.

\section{Sources with time-dependent neutrino emission profiles}
\label{sec:time-dep}

\noindent The results shown in this contribution assume steady neutrino emission as done in Ref.~\cite{PhysRevLett.124.051103}.
Besides classes for spatial and energy PDFs, SkyLLH also provides a class for time PDFs. Hence, the signal PDF in Eq.~(\ref{eq:signal-pdf}) can be modified to include a time PDF:
\begin{equation}
    \mathcal{S}(\delta_i, E_i|\delta_{\rm{src}},\gamma) = P_\mathrm{PSF}(\delta_i|\delta_{\mathrm{src}}) \cdot P_{\rm{E}}(E_i|\delta_{\mathrm{src}},\gamma) \cdot P_{\rm{T}}(t_i|t_0,\sigma_{\rm{T}}).
    \label{eq:signal-pdf-timedep}
\end{equation}
The current version includes a box-shaped time PDF (assuming constant emission between a starting and end time) and a Gaussian-shaped time PDF centered at $t_0$ with width $\sigma_{\rm{T}}$. Using the box-shaped time profile from Ref. \cite{IceCube:2021xar} with start and stop times, respectively, of $t_{\mathrm{start}}=56927.86$~(MJD) and $t_{\mathrm{stop}}=57116.76$~(MJD), the best-fit parameters for the flare of the blazar TXS 0506+056 are $\hat{n}_{\rm{s}} = 11.66$ and $\hat{\gamma}=2.25$, agree with values published in Ref.~\cite{IceCube:2021xar}, namely $\hat{n}_{\rm{s}} = 11.87$ and $\hat{\gamma}=2.22$.

\noindent SkyLLH supports the fitting of the parameters of a Gaussian-shaped time PDF (i.e., $t_0, \sigma_{\rm{T}}$) using the unsupervised learning algorithm ``Expectation Maximization (EM)'' \cite{Dempster-EM-1977}.
EM is applied to the time sequence of the events, where each event is weighted with its signal-over-background PDF ratio of the spatial and energy PDFs, as in Eq.~(\ref{eq:ts-llh-ratio}). With the fitted parameters, we create the time PDF and subsequently optimize the likelihood function. Since the PDF ratio values depend on $\gamma$, we repeat these steps for $\gamma$ values ranging from 1 to 5 and take the maximal test-statistic from all EM fits as the final test-statistic value. The EM method was used in Ref.~\cite{martina_alert:2023icrc} with IceCube internal data where the fitted flare parameter and the p-value for the flare of TXS~0506+056 are comparable with published results in Ref.~\cite{doi:10.1126/science.aat2890, IceCube:2021xar, IC-2021-10year-multi-flare}. With the published data set we estimate the significance of the observed flare (using EM) and find a local p-value of $-\log_{10}(p_{\rm{loc}}) = 1.23$. The reduced p-value compared to previous works is expected, see also Section \ref{sec:reproducibility}, Figure \ref{fig:sensitivity-comparison}. The best-fit parameters for the time-dependent neutrino emission from TXS~0506+056 using the public data are $\hat{n}_{\rm{s}} = 7.58$ and $\hat{\gamma} = 2.21$, centered at $\hat{t}_0 = 56972.65$~(MJD) with $\hat{\sigma}_{\rm{T}}= 27.97$~days. In Ref.~\cite{IC-2021-10year-multi-flare}, the flare parameters were $\hat{t}_0 = 57000 \pm 30 $~(MJD), $\hat{\sigma}_{\rm{T}} = 62 \pm 27$~days, $\hat{n}_{\rm{s}} = 10^{+5.2}_{-4.2}$, and $\hat{\gamma} = 2.2 \pm 0.3$. However, it should be noted that the fitting and analysis procedures in Ref.~\cite{IC-2021-10year-multi-flare} are different, specifically concerning the signal energy PDF (see Section \ref{sec:lh-formalism}) and, therefore, the signal-over-background PDF-ratio weights.

\section{Conclusion and outlook}
\label{sec:conclusion}

\noindent In this contribution, we demonstrated an extension of SkyLLH to perform time-integrated and time-dependent neutrino point-source searches based on 10-year public point-source data. By using this extension, published IceCube results \cite{PhysRevLett.124.051103} can be reproduced reasonably well.
We plan to implement additional features to support general source flux functions and the ability to evaluate the cumulated signal from multiple sources through point-source stacking analyses in the future.  

\bibliographystyle{ICRC}
\bibliography{references}

%

\clearpage

\section*{Full Author List: IceCube Collaboration}

\scriptsize
\noindent
R. Abbasi$^{17}$,
M. Ackermann$^{63}$,
J. Adams$^{18}$,
S. K. Agarwalla$^{40,\: 64}$,
J. A. Aguilar$^{12}$,
M. Ahlers$^{22}$,
J.M. Alameddine$^{23}$,
N. M. Amin$^{44}$,
K. Andeen$^{42}$,
G. Anton$^{26}$
% ...
\\
\\
% ...
$^{12}$ Universit{\'e} Libre de Bruxelles, Science Faculty CP230, B-1050 Brussels, Belgium \\
% ...
$^{17}$ Department of Physics, Loyola University Chicago, Chicago, IL 60660, USA \\
$^{18}$ Dept. of Physics and Astronomy, University of Canterbury, Private Bag 4800, Christchurch, New Zealand \\
% ...
$^{22}$ Niels Bohr Institute, University of Copenhagen, DK-2100 Copenhagen, Denmark \\
$^{23}$ Dept. of Physics, TU Dortmund University, D-44221 Dortmund, Germany \\
% ...
$^{26}$ Erlangen Centre for Astroparticle Physics, Friedrich-Alexander-Universit{\"a}t Erlangen-N{\"u}rnberg, D-91058 Erlangen, Germany \\
% ...
$^{40}$ Dept. of Physics and Wisconsin IceCube Particle Astrophysics Center, University of Wisconsin{\textendash}Madison, Madison, WI 53706, USA \\
% ...
$^{42}$ Department of Physics, Marquette University, Milwaukee, WI, 53201, USA \\
% ...
$^{44}$ Bartol Research Institute and Dept. of Physics and Astronomy, University of Delaware, Newark, DE 19716, USA \\
% ...
$^{63}$ Deutsches Elektronen-Synchrotron DESY, Platanenallee 6, 15738 Zeuthen, Germany  \\
$^{64}$ Institute of Physics, Sachivalaya Marg, Sainik School Post, Bhubaneswar 751005, India \\
% ...

\subsection*{Acknowledgements}

\noindent
The authors gratefully acknowledge the support from the following agencies and institutions:
USA {\textendash} U.S. National Science Foundation-Office of Polar Programs,
U.S. National Science Foundation-Physics Division,
U.S. National Science Foundation-EPSCoR,
Wisconsin Alumni Research Foundation,
Center for High Throughput Computing (CHTC) at the University of Wisconsin{\textendash}Madison,
Open Science Grid (OSG),
Advanced Cyberinfrastructure Coordination Ecosystem: Services {\&} Support (ACCESS),
Frontera computing project at the Texas Advanced Computing Center,
U.S. Department of Energy-National Energy Research Scientific Computing Center,
Particle astrophysics research computing center at the University of Maryland,
Institute for Cyber-Enabled Research at Michigan State University,
and Astroparticle physics computational facility at Marquette University;
Belgium {\textendash} Funds for Scientific Research (FRS-FNRS and FWO),
FWO Odysseus and Big Science programmes,
and Belgian Federal Science Policy Office (Belspo);
Germany {\textendash} Bundesministerium f{\"u}r Bildung und Forschung (BMBF),
Deutsche Forschungsgemeinschaft (DFG),
Helmholtz Alliance for Astroparticle Physics (HAP),
Initiative and Networking Fund of the Helmholtz Association,
Deutsches Elektronen Synchrotron (DESY),
and High Performance Computing cluster of the RWTH Aachen;
Sweden {\textendash} Swedish Research Council,
Swedish Polar Research Secretariat,
Swedish National Infrastructure for Computing (SNIC),
and Knut and Alice Wallenberg Foundation;
European Union {\textendash} EGI Advanced Computing for research;
Australia {\textendash} Australian Research Council;
Canada {\textendash} Natural Sciences and Engineering Research Council of Canada,
Calcul Qu{\'e}bec, Compute Ontario, Canada Foundation for Innovation, WestGrid, and Compute Canada;
Denmark {\textendash} Villum Fonden, Carlsberg Foundation, and European Commission;
New Zealand {\textendash} Marsden Fund;
Japan {\textendash} Japan Society for Promotion of Science (JSPS)
and Institute for Global Prominent Research (IGPR) of Chiba University;
Korea {\textendash} National Research Foundation of Korea (NRF);
Switzerland {\textendash} Swiss National Science Foundation (SNSF);
United Kingdom {\textendash} Department of Physics, University of Oxford.

\end{document}